\documentclass[natbib]{emulateapj}
\usepackage{natbib}
\usepackage{graphicx}
\usepackage{bm}
\input epsf
\def\s2n{S^{\prime}/N}

\def\gsim{\;\rlap{\lower 2.5pt
\hbox{$\sim$}}\raise 1.5pt\hbox{$>$}\;}
\def\lsim{\;\rlap{\lower 2.5pt
\hbox{$\sim$}}\raise 1.5pt\hbox{$<$}\;}

\usepackage{amssymb,amsmath}
\usepackage{mathrsfs}
\def\bs{\boldsymbol}

\begin{document}
\title{The Probability Distribution of Density Fluctuations in Supersonic Turbulence }

\author{Liubin Pan$^1$}
\author{Paolo Padoan$^{2,3}$}
\author{{\AA}ke Nordlund$^4$}
\affiliation{$^1$School of Physics and Astronomy, Sun Yat-sen University, 2 Daxue Road, Zhuhai, Guangdong, 519082, China; panlb5@mail.sysu.edu.cn}
\affiliation{$^2$Institut de Ci\`{e}ncies del Cosmos, Universitat de Barcelona, IEEC-UB, Mart\'{i} Franqu\`{e}s 1, E08028 Barcelona, Spain; 
ppadoan@icc.ub.edu}
\affiliation{$^3$ICREA, Pg. Llu\'{i}s Companys 23, 08010 Barcelona, Spain}
\affiliation{$^4$Centre for Star and Planet Formation, Niels Bohr Institute and Natural History Museum of Denmark, University of Copenhagen, {\O}ster Voldgade 5-7, DK-1350 Copenhagen K, Denmark; aake@nbi.ku.dk}

\begin{abstract}
We study density fluctuations in supersonic turbulence using both theoretical methods and numerical simulations. 
A theoretical formulation is developed for the probability distribution function (PDF) of the density at steady state, connecting it to the conditional 
statistics of the velocity divergence. Two sets of numerical simulations are carried out, using either a Riemann solver to evolve the Euler equations 
or a finite-difference method to evolve the Navier-Stokes (N-S) equations. After confirming the validity of our theoretical formulation with the N-S simulations, 
we examine the effects of dynamical processes on the PDF, showing that the nonlinear term in the divergence 
equation amplifies the right tail of the PDF and reduces the left one, the pressure term reduces both the right and left tails, and the viscosity term, 
counter-intuitively, broadens the right tail of the PDF. Despite the inaccuracy of the velocity divergence from the Riemann runs, as found in our 
previous work, we show that the density PDF from the Riemann runs is consistent with that from the N-S runs. Taking advantage of their much higher effective 
resolution, we then use the Riemann runs to study the dependence of the PDF on the Mach number, $\mathcal{M}$, up to $\mathcal{M}\sim30$. The PDF width, 
$\sigma_{s}$, follows the relation $\sigma_{s}^2 = \ln (1+b^2 {\mathcal M}^2)$, with $b\approx0.38$. However, the PDF exhibits a negative skewness that 
increases with increasing $\mathcal{M}$, so much of the growth of $\sigma_{s}$ is accounted for by the growth of the left PDF tail, while the growth of the right 
tail tends to saturate. Thus, the usual prescription that combines a lognormal shape with the standard variance-Mach number relation greatly overestimates the 
right PDF tail at large $\mathcal{M}$, which may have a significant impact on theoretical models of star formation. 
\end{abstract}

\maketitle

\section{Introduction}

The probability distribution function (PDF) of density fluctuations in supersonic turbulence is of crucial importance for 
understanding and modeling star formation in molecular clouds (MCs).  Numerical studies have established that the 
density PDF of isothermal, supersonic turbulence with solenoidal driving is generally well described by a lognormal distribution 
\citep{Vazquez-Semadeni94,Padoan+97imf,Nordlund+Padoan99pdf,Ostriker+01,Kritsuk+07,Federrath+08,Lemaster+Stone08,Federrath+10}, 
although departures from lognormality have also been found in flows at high Mach numbers and/or with compressive 
driving \citep{Federrath+10,Federrath13} and in flows with self-gravity \citep[e.g.][]{Collins+11,Kritsuk+11pdf,Collins+12}. 
The lognormal PDF has been widely used in star-formation models based on 
turbulent fragmentation to predict the star formation rate \citep{Krumholz+McKee05,Padoan+Nordlund11sfr,Federrath+Klessen12} 
and the stellar initial mass function \citep{Padoan+97imf,Padoan+Nordlund02imf,Hennebelle+Chabrier08,Hennebelle+Chabrier11}. 

Despite its popularity, the physical origin of the lognormal density PDF remains poorly understood. Attempts to explain the lognormal
PDF have relied on phenomenological arguments, such as the idea that it originates from a multiplicative process of independent 
compression and expansion events \citep{Vazquez-Semadeni94}, or on qualitative argument based on an exact PDF 
equation of \citep{Pope+Ching93} without specifying the physics behind the terms in the equation 
\citep{Nordlund+Padoan99pdf}. Even the observational evidence of a lognormal PDF in MCs is 
questionable. The observational studies measure the column density PDF, but a lognormal 
(or power-law) column density PDF does not necessarily imply a lognormal (or power-law) density PDF, and vice versa. Furthermore,
early results showing that the column density PDF in MCs is lognormal if self-gravity is not dominant \citep{Kainulainen+09,Kainulainen+14} 
have been later interpreted as the effect of incomplete sampling, with the intrinsic PDF being consistent with a power law \citep{Lombardi+15,Alves+17};
and even power-law tails may be artifacts of observational biases \citep{Brunt15}. 

In this work, we examine a number of important issues concerning the density PDF in supersonic turbulence.  
First, we make an attempt to understand what determines the shape of the density PDF from first principles. 
Rather than using phenomenological arguments, we will derive exact equations for the density PDF from the continuity 
and Navier-Stokes equations. Combined with numerical simulations, the equations will help us understand the 
physics behind the density PDF. In particular, we will investigate the effects of the dynamical processes on the PDF 
shape.  

The second issue to be explored in this study is the reliability of the density PDF measured from numerical
simulations using Riemann solvers. In our recent work (Pan et al. 2019), we have shown that spatial derivatives
based on cell-center values computed from solutions of turbulent flows based on Riemann solvers are inaccurate, 
particularly in regions with flow discontinuities, while they are very accurate in the case of the smoother, 
direct solutions of the Navier-Stokes equation with explicit viscosity.
A consequence of using inaccurate derivatives is that the continuity equation appears to be violated,
and the simulation data fail the tests against a number of exact results derived from the continuity
equation involving the spatial derivatives in the flow. This might raise concerns about the accuracy 
of the density field in the Riemann runs. 
Note, however, that the inaccurate spatial derivatives are not directly used in the finite-volume 
method that advances the solution, and that approximate Riemann solver methods (such as the HLLC method
employed here) produce results that closely approximate solutions of the Euler equation.

Here, we will test the accuracy of the density PDFs from Riemann-solver simulations by comparison
with the PDFs from Navier-Stokes simulations with explicit viscosity. This is a crucial test because
at high Mach numbers we must rely on Riemann runs, as the numerical convergence of the density PDF
requires a relatively high Reynolds number that cannot be achieved (at a reasonable computational cost)
by the simulations with explicit viscosity.

Finally, we will investigate the Mach number dependence of the density PDF using Riemann-solver simulations with Mach numbers up to $\simeq 30$. Theoretical models 
of star formation based on turbulent fragmentation rely on the extrapolation of the density PDF at relatively low Mach 
numbers, usually adopting a lognormal distribution for the shape of the density PDF, in combination 
with a variance-Mach number relation for the PDF width \citep{Padoan+97imf,Nolan+15}. The  
application of such a prescription to environments with very high Mach numbers is questionable, because 
at large Mach numbers the density PDF shows significant negative skewness, even in the case of 
solenoidal driving \citep{Federrath13}. Examples of such dubious extrapolations of the standard 
lognormal model to very large Mach numbers include the prediction of a bottom-heavy IMFs in progenitors 
of early-type galaxies and extreme starburst environments \citep{Hopkins13,Chabrier+14}. In this
work, we will show strong deviations of the high-density tail of the PDF relative to those commonly-used assumptions. 


In \S 2, we provide a theoretical formulation for the density PDF in supersonic turbulence based on the 
derivation of an exact PDF equation. \S 3 describes the numerical method and the simulation 
setup. Two sets of simulations are carried out, one based on a Riemann solver to evolve the Euler 
equation, and the other based on a staggered mesh code that solves the Navier-Stokes (N-S) equation with explicit viscosity. 
\S 4 verifies our theoretical formulation using the data from N-S simulations. The formulation is then applied to investigate 
the physics behind the density PDF in supersonic turbulence. In order to assess the accuracy of the density PDF from 
Riemann-solver simulations, \S 5 compares the results from the two sets of simulations. In \S 6, we examine the dependence 
of the density PDF on the Mach number, and our conclusions are summarized in \S 7. 

\section{Theoretical Formulation for the Density PDF}

Our theoretical formulation for the density PDF in supersonic turbulence was motivated partly 
by the work of \citet{Pope+Ching93}, who derived a general formula for the PDF of any quantity of interest 
in statistically stationary systems. The formula shows that at steady state the PDF of a quantity of interest can be 
expressed in terms of the conditional expectations of its time derivatives (see Appendix A for a derivation of the 
formula). The formula of \citet{Pope+Ching93} was applied by \citet{Nordlund+Padoan99pdf} to interpret the 
density PDF in supersonic turbulence. 
However, the work of \citet{Pope+Ching93} was originally developed for the study of passive scalars, 
not specifically for density fluctuations in compressible turbulence. In fact, their formulation does not make use of the 
hydrodynamic equations, and thus may not reveal the physics behind the density PDF in supersonic turbulence,
unless some physical insight about the conditional expectations of the time derivatives of the density is obtained
through the momentum and continuity equations \citep{Nordlund+Padoan99pdf}. 

In this work, we will use the probabilistic approach to study density fluctuations in supersonic turbulence.  
We develop a theoretical formulation for the density PDF using the continuity equation, 
\begin{equation}
\frac {\partial \rho }{\partial t}  + \nabla \cdot  (\rho \bs{u} )  =  0, 
\label{rho-eq}
\end{equation}
and the Navier-Stokes equation, 
\begin{equation}
\frac {\partial  \bs{u} }{\partial t}  +     \bs{u}  \cdot  \nabla \bs{u} = - \frac{\nabla p}{\rho} + \frac{1}{\rho} \nabla \cdot \mathcal{\sigma}   + {\bs a},
\label{v-eq}
\end{equation}  
where ${\bs a}$ is the acceleration that drives the turbulent flow, and $\sigma$ is the viscous tensor.  All other symbols carry their conventional 
meanings.  Based on the flow equations, our approach provides a more direct physical understanding of the PDF than the formula of 
\citet{Pope+Ching93}. We will first derive an exact equation for the density PDF from the continuity equation. At steady state, the equation leads to a 
formula for the PDF, relating it to the conditional statistics of the flow divergence. The formula will then be combined 
with the Navier-Stokes equation to study the effects of the dynamic processes on the PDF shape. 




\subsection{A formula for the density PDF at steady state}

We start by deriving an exact equation for the density PDF in compressible turbulence.  
For convenience, we rewrite the continuity equation as,    
\begin{equation}
\frac {\partial s}{\partial t}  + \bs{u}  \cdot \nabla s = - \nabla \cdot \bs{u},
\label{s-eq}
\end{equation}
where $s\equiv \ln (\rho/\bar{\rho})$ is the logarithm of the density, $\rho$, with $\bar{\rho}$ the mean 
density. Following the standard procedure of the probabilistic approach to turbulence \citep[e.g.][]{Pope00},   
we first define a fine-grained PDF of $s$ as $g(\zeta;  {\bs x}, t) = \delta(\zeta - s({\bs x}, t))$,  
where $\delta$ is the Dirac delta function and $\zeta$ the sampling variable. 
Since $g$ depends on $t$ and ${\bs x}$ only through  $\zeta - s({\bs x}, t)$, 
the time derivative and spatial gradient of $g$ are given by $\partial_t g (\zeta; {\bs x}, t) = -\partial_\zeta g \partial_t s $ and $\nabla g = - \partial_\zeta g \nabla s$, respectively (Pan et al.\ 2018). 
Using Eq.\ (\ref{s-eq}), we find that, 
 \begin{equation}
  \frac {\partial g (\zeta; {\bs x}, t)}{\partial t} +     \bs{u}  \cdot  \nabla g  = 
   \frac{ \partial (g \nabla \cdot \bs{u}) }{\partial \zeta }. 
\label{fine2}
\end{equation}
where the last term uses the fact that $\nabla \cdot \bs{u}$ is independent of the sampling variable, $\zeta$.


Using Eq.\ (\ref{fine2}), it is straightforward to show that,  
 \begin{flalign}
 \frac {\partial (g \nabla \cdot {\bs u} )}{\partial t}  +  \nabla  \cdot   ( g \bs{u} &\nabla \cdot {\bs u})      =    g \frac{d (\nabla \cdot {\bs u}) } {dt}  \nonumber\\
                                                                                   & 
                                                                + \exp(-\zeta )  \frac{ \partial [\exp(\zeta )g (\nabla \cdot \bs{u} )^2 ] }{\partial \zeta }                                                                                                  
\label{intermediate}
\end{flalign}
where $d (\nabla \cdot {\bs u})/dt =  \partial_t (\nabla \cdot {\bs u}) + \bs{u} \cdot \nabla (\nabla \cdot {\bs u})$ denotes the Lagrangian time-derivative of the flow divergence.

For any flow quantity, $\psi({\bs x}, t)$, we define its average as $\langle \psi \rangle = \frac{1}{V}\int_V  \psi({\bs x}, t) d^3x$ where $V$ is the volume of the flow domain.
We then define the coarse-grained PDF as the average of the fine grained PDF: $f(\zeta; t) \equiv \langle g(\zeta; {\bs x}, t) \rangle =   \langle \delta(\zeta - s({\bs x}, t)) \rangle$.
The conditional mean, $\langle  \psi ({\bs x}, t)| s({\bs x}, t)=\zeta\rangle$, of a flow quantity, $\psi$, is the 
average of $\psi$ over all the flow regions where $s( {\bs x}, t)$ equals the sampling variable, and it 
can be computed as $ \langle  \psi ({\bs x}, t)| s({\bs x}, t)= \zeta \rangle  = \int_V  \psi ({\bs x}, t) g(\zeta; {\bs x}, t) d^3x/ \int_V g(\zeta; {\bs x}, t) d^3x$,  which is equal to 
$\langle \psi ({\bs x}, t) g(\zeta; {\bs x}, t) \rangle/f(\zeta; t)$.
Taking the average of Eq.\ (\ref{intermediate}) and using the definition of the conditional mean leads to, 
\begin{flalign}
  \frac {\partial  ( f \langle \nabla \cdot {\bs u}| s=\zeta\rangle)}{\partial t}   =  &   \exp(-\zeta) \frac{ \partial [   \exp(\zeta) f \langle (\nabla \cdot \bs{u})^2|s=\zeta\rangle]}{\partial \zeta } \nonumber \\&+ 
   f  \left\langle \frac{d (\nabla \cdot \bs{u})}{dt} \bigg|s =\zeta\right\rangle,
\label{final}
\end{flalign}
where we used the fact that the second term in Equation (\ref{intermediate}) is a surface term, and thus vanishes either from the periodic boundary condition 
or equivalently from the assumption of statistical homogeneity. 
For simplicity of notations,  we will drop the sampling variable, $\zeta$, and write $\langle ...|s =\zeta\rangle $ simply as $\langle ...|s\rangle $. The density PDF will 
be written as $f(s, t)$, accordingly. 

At steady state, the time derivative term in Eq.\ (\ref{final}) vanishes
and one obtains a formal solution for the PDF of $s$,  
\begin{equation}
f(s) = C \exp(-s) D^{-1}(s) \exp\left[ \int\limits_0^{s}  \frac  {N (s')}{D(s')} ds' \right],
\label{solution}
\end{equation}
where $C$ is the integration constant to be fixed by the normalization of the PDF.  The $D$ and $N$ terms in the equation are 
the conditional averages in Equation (\ref{final}) normalized to the variance of the flow divergence,  i.e., $D \equiv \langle (\nabla \cdot \bs{u})^2|s\rangle/\sigma_{\rm d}^2$ 
and $N\equiv -\langle d (\nabla \cdot \bs{u})/dt |s\rangle/\sigma_{\rm d}^2$, where 
$\sigma_{\rm d} = \langle (\nabla \cdot {\bs u})^2\rangle^{1/2}$ is the root-mean-square of the divergence. 

Equation (\ref{solution}) may be derived in different ways, and one of the alternative derivations 
is given in Appendix B.   The derivation in Appendix B applies the method of \citet{Pope+Ching93} to 
the density PDF defined in the Lagrangian frame, which is then converted to  
the Eulerian frame using the relation between Eulerian and Lagrangian PDFs. 


\subsection{The dynamical effects on the density PDF}

Equation (\ref{solution}) can be used to study the effects of various dynamical effects on the density PDF. 
The dynamical effects enter the formulation through the $N$ term, 
which depends on the evolution of the velocity divergence. 
Taking the divergence of the Navier-Stokes Equation (\ref{v-eq}) gives, 
\begin{flalign}
\frac{d { (\nabla \cdot \bs u})}{d t}  = & - \nabla \bs u :  \nabla \bs u -\nabla \cdot \left(\frac {\nabla p} {\rho}\right) \nonumber \\&  + \nabla \cdot \left(\frac{1}{\rho}\nabla\cdot\sigma\right)+\nabla \cdot {\bs a}, 
\label{div-eq}
\end{flalign}
where the nonlinear term $\nabla \bs u :  \nabla \bs u \equiv \partial_i u_j  \partial_j u_i$ is the trace of $\nabla \bs u \cdot \nabla \bs u$. 
Intuitively, in converging dense regions, the nonlinear term tends to self-amplify 
the amplitude of the divergence, while the gas pressure resists compressions, and reduces the divergence amplitude. 

It follows immediately from the definition of $N$($\equiv -\langle d (\nabla \cdot \bs{u})/dt |s\rangle/\sigma_{\rm d}^2$) and Eq.\ (\ref{div-eq}) that,
\begin{equation}
N =  N_{\rm nl}  + N_{\rm p}  + N_{\nu} + N_{\rm a},
\label{N-eq}
\end{equation}
where $N_{\rm nl} = \langle \nabla \bs u :  \nabla \bs u |s \rangle/\sigma_{\rm d}^2$, $N_{\rm p} =  \langle \nabla\cdot(\rho^{-1}\nabla p) |s \rangle /\sigma_{\rm d}^2$, 
$N_{\nu}=-\langle\nabla \cdot (\rho^{-1} \nabla \cdot \sigma )|s\rangle/\sigma_{\rm d}^2$
and $N_{\rm a} =-\langle  \nabla \cdot {\bs a}|s  \rangle/\sigma_{\rm d}^2$.  All the $N$ terms are normalized to the variance of the divergence, 
$\sigma_{\rm d}^2 = \langle (\nabla \cdot {\bs u})^2\rangle$. The four terms in Equation (\ref{N-eq}) represent the effects of four dynamical processes, 
namely, the nonlinear effect, the pressure, the viscosity and the driving force, respectively. For an isothermal gas,  
$N_{\rm p} =  C_{\rm s}^2 \langle \nabla^2 s|s \rangle /\sigma_{\rm d}^2$ with $C_{\rm s}$ the constant sound speed. 

If the driving acceleration is solenoidal, we have $N_{\rm a} =0$, and the density PDF is given by,
\begin{flalign}
f(s) = & C \exp(-s) D^{-1}(s) \times  \nonumber\\ 
&\exp\left[ \int\limits_0^{s}  \frac {N_{\rm nl} (s') + N_{\rm p} (s') +N_{\nu} (s')}{D(s')} ds' \right].
\label{solution2}
\end{flalign}
Combined with simulation data, this equation will help us understand the effects of various dynamical 
processes on the density PDF in compressible turbulence.

\section{Numerical Simulations}

One of the goals of the current study is to understand the physics behind the density PDF in supersonic turbulence by analyzing the simulation 
data according to our theoretical formulation, i.e., Eqs.\ (\ref{solution}) and (\ref{solution2}).  Both equations depend on the conditional expectations of 
the spatial gradients of the density and/or velocity fields. Thus, the verification or application of the formulation requires 
numerical simulations that provide accurate estimates for the spatial derivatives. Recently, Pan et al.\ (2019) have 
shown that spatial derivatives computed from the post-processing of simulations of supersonic turbulence using 
Riemann solvers without explicit viscosity are inaccurate, especially in regions with large density and near-discontinuities. 
Therefore, these simulations, which we refer to as 
Riemann simulations, are not suitable for the study of density fluctuations in supersonic turbulence with the 
theoretical formulation presented in \S 2.  

The spatial gradients can be accurately evaluated in simulations that 
evolve the Navier-Stokes equation (\ref{v-eq}) with the viscous term 
explicitly included (Pan et al.\ 2019).  Following Pan et al.\ (2019), 
we name simulations with explicit viscosity the N-S simulations. 
The N-S simulations make it possible to investigate the density PDF using Equations (\ref{solution}) and (\ref{solution2}). 
In particular, the N-S simulations allow a direct computation of the $N_{\nu}$ term in Eq.\ (\ref{solution2}). 
This is in contrast to the Riemann simulations where the viscosity is implicit through 
numerical diffusion so that $N_{\nu}$ cannot be directly evaluated.  


Strictly speaking, simulations of supersonic interstellar turbulence should include the physical 
viscosity, as it exists in the interstellar gas. However, including explicit viscosity leads to 
a small Reynolds number and thus a very short, if any, inertial range even at a resolution of 1024$^3$.  
The N-S simulations have the potential problem of not being able to resolve strong density fluctuations, 
especially at large Mach numbers. The Riemann simulations, on the other hand, may achieve 
significantly higher effective resolution. In Riemann simulations, the dissipation of 
kinetic energy occurs through the numerical diffusion only, and the simulated flow may acquire the largest effective 
Reynolds number hence the broadest inertial range possible at a given numerical resolution.


Considering their strengths and weaknesses, we conducted both N-S and Riemann simulations. 
By comparing the two sets of simulation runs at similar Mach numbers, we will check the 
reliability and accuracy of the density PDFs measured in the N-S and Riemann simulations.  
Both sets of simulations were carried out with the \emph{Dispatch} code \citep{Nordlund+18}, which provides 
an efficient computing framework to test different fluid-dynamics solvers.  For the Riemann runs, 
we adopted the HLLC (Harten-Lax-van Leer-Contact) approximate Riemann solver \citep{Toro+94}. 
The N-S runs used a simplified, 2nd order version of the 6th order solver of the \emph{Stagger Code} \citep{Galsgaard+96,Kritsuk+11,Baumann+12}, 
including the viscous term in the Navier-Stokes equation (\ref{v-eq}).
The viscous tensor, $\sigma$, in the N-S runs was set to $\sigma_{ij} = \rho \nu (\partial_j u_i + \partial_i u_j)$. 
A discussion about this choice for $\sigma_{ij}$ can be found in Pan et al.\ (2019).
The kinematic viscosity was set to $\nu = \Delta x \,\mathcal{M} C_{\rm s}$, where the speed of sound $C_{\rm s}$ 
is unity in all simulations. The linear scaling of $\nu$ with the computational cell size, $\Delta x$, 
and the Mach number  $\mathcal{M}$ is sufficient to maintain numerical stability in all flows. 
With such a scaling, the Reynolds number is invariant with $\mathcal{M}$, and increases linearly 
with the numerical resolution, being roughly $\simeq 1024$ in a N-S run with $1024^3$ computational elements. 

For all simulations, we adopted an isothermal equation of state, and evolved the hydrodynamic equations 
in a three-dimensional (3D) simulation box with periodic boundary conditions. The simulated flows were driven and maintained by 
a random, solenoidal force generated in Fourier space using wave numbers in the range $1<kL/2\pi<2$, where $L$ is the box 
size. We conducted N-S simulations at three Mach numbers, $\mathcal{M} = 2.1$, $3.8$ and $7.2$. 
Due to the limited numerical resolution and the inclusion of an explicit viscous term, the N-S simulations may not sufficiently 
resolve density fluctuations at higher Mach numbers. The Riemann runs, on the other hand, have significantly larger effective 
numerical resolution. Thus, in addition to the same three Mach numbers as in the N-S case,
we also experimented with three higher Mach numbers, $\mathcal{M} =11.6, 15.2$ and $29.8$. 
The highest numerical resolution was 1024$^3$ for both N-S and Riemann runs, and, for the study of numerical convergence 
or the Reynolds number dependence, we have also conducted some runs at lower resolutions, 256$^3$ and $512^3$.  

All simulations are integrated for two sound crossing times, i.e., $2L/C_{\rm s}$. 
In units of the dynamical time, $T$, defined as $L/2U$, with $U$ the 3D rms velocity of the flow, each simulations is integrated for $2\mathcal{M} T$. 
We saved 100 snapshots per simulation, equally-spaced in time, but only used the last 81 snapshots in the analysis, 
to avoid initial transients as the flow evolves from the initial conditions (uniform density and zero velocity),
and focus on the steady-state statistics.


\begin{figure*}[t]
\includegraphics[width=2\columnwidth]{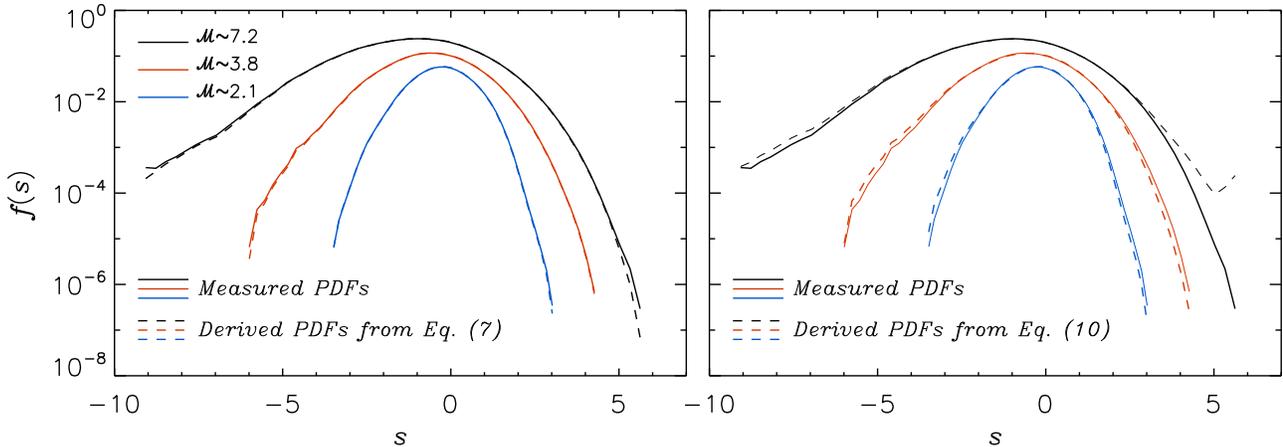}
\caption[]{Density PDFs from 1024$^3$ N-S runs at  three Mach numbers $\mathcal{M}=2.1$, 3.8, and 7.2.  The solid lines 
show the PDFs directly measured from the simulation data.  The dashed lines in the left panel plot the derived PDFs 
from Eq.\ (\ref{solution}) using the conditional statistics of the velocity divergence. In the right panel, the dashed lines are 
calculated from Eq.\ (\ref{solution2}) using the three dynamical contributions, $N_{\rm nl}$, $N_{\rm p}$ and $N_{\nu}$, computed from the data. 
For clarity, the PDFs at different Mach numbers are shifted vertically. }
\label{pdfs}
\end{figure*}

\section{Understanding the density PDF in supersonic turbulence}

\subsection{Verification of the theoretical formulation} 

The formulation developed in \S 2 allows us to shed some light on the physics behind the PDF 
of density in supersonic turbulence. We start by verifying the formulas derived in \S 2 using the N-S 
simulation data. Fig.\ \ref{pdfs} shows the density PDF from $1024^3$ N-S runs at $\mathcal{M}=2.1$, 3.8 
and 7.2. In both panels, the solid lines plot the density PDF directly measured from the simulation data. 
The PDF of $s$ at Mach 2.1 is approximately symmetric, close to a lognormal distribution, but 
as $\mathcal{M}$ increases, the left tail broadens faster than the right one, and the PDF shows 
stronger negative skewness.  

In the left panel, the dashed lines correspond to the PDFs calculated from 
Equation (\ref{solution}) using $D$ and $N$ computed from the simulation data. 
The Lagrangian time-derivative of divergence in $N(s)$ was evaluated as $d(\nabla \cdot {\bs u})/dt=  \nabla \cdot (\partial_t {\bs u}) + \bs u \cdot  \nabla (\nabla \cdot {\bs u})$.
The partial time derivative of the velocity was obtained by calculating the 
velocity changes at neighboring time steps during the simulation run.  The left panel shows that, 
for all Mach numbers, the dashed lines match very well the solid lines, confirming the validity of our 
formula, Equation (\ref{solution}), for the density PDF at steady state. Since 
our derivation of the PDF formula  is exact, the figure actually demonstrates that the first and second 
order velocity derivatives, needed  for the calculation of $D$ and $D$,  are accurately 
evaluated in the N-S simulations. 
 
The dashed lines in the right panel plot the PDFs calculated from Equation (\ref{solution2}) in \S 2.2. 
The nonlinear ($N_{\rm nl}$), pressure ($N_{\rm p}$) and viscosity ($N_{\nu}$) terms in that equation are 
computed from the simulation data according to their definitions. The derived PDFs in the right panel also agree 
well with the measured ones, except at the right tail of the PDF for $\mathcal{M}=7.2$. 
Equations (\ref{solution}) and (\ref{solution2}) are equivalent to each other, as they are connected by 
Equation (\ref{N-eq}),  i.e., $N= N_{\rm nl} + N_{\rm p} + N_{\nu}$ for solenoidal driving, which follows from the Navier-Stokes equation. Therefore, the derived PDFs from Equations (\ref{solution}) and (\ref{solution2}) 
are expected to be identical and both agree with the measured PDFs. 

However, numerical errors may enter the computation of the $N$ terms, which involve high-order spatial derivatives of the velocity and density fields. 
For example, the evaluation of $N_{\rm p}$ requires second order derivatives of $s$, and the computation 
of the viscous term, $N_{\nu}$, involves both third order derivatives of velocity and second order derivatives of $s$.  
The discrepancy at the right PDF tail for $\mathcal{M} =7.2$ in the right panel indicates 
numerical errors in the evaluation of the $N$ terms in Equation (\ref{solution2}).  
The errors probably come from the third order spatial derivatives in $N_{\nu}$, as 
the accuracy of higher order derivatives is more difficult to achieve. 
On the other hand, since the discrepancy occurs only at the largest $s$ for the largest Mach number, 
it may also be caused by the inaccuracy of the second order derivatives of $s$ at large densities, 
which exist in both $N_{\rm p}$ and $N_{\nu}$.  Note that the evaluation of $N_{\rm nl}$, 
which only depends on the first order derivatives of velocity, is likely accurate. 
We find that the discrepancy at the right tail for the $\mathcal{M} =7.2$ flow decreases  
when a larger value for the viscosity, $\nu$, is adopted.  
With a larger $\nu$, the simulated flow is smoother, and the evaluation 
for the high order spatial derivatives becomes more accurate.  


In our N-S simulations,  the Reynolds number increases linearly with the numerical resolution. 
By comparing with results from lower resolutions, we found that for the two lower Mach 
numbers, the density PDF has reached convergence at $Re \simeq 1000$. 
On the other hand, in the Mach 7.2 case, the right tail of the PDF is still broadening with the 
Reynolds number at $Re \simeq 1000$, and simulations at higher resolutions and larger 
Reynolds numbers are required to achieve the convergence of the PDF for $ \mathcal{M} \gsim 7$.  

In summary, the general agreement between the derived PDFs from our formulas, Equations (\ref{solution}) and (\ref{solution2}), 
and the directly measured PDFs confirms the validity of our theoretical formulation from \S 2, which may be used 
in combination with our N-S simulations to study the physics behind the density PDF. 

\begin{figure*}
\includegraphics[width=2\columnwidth]{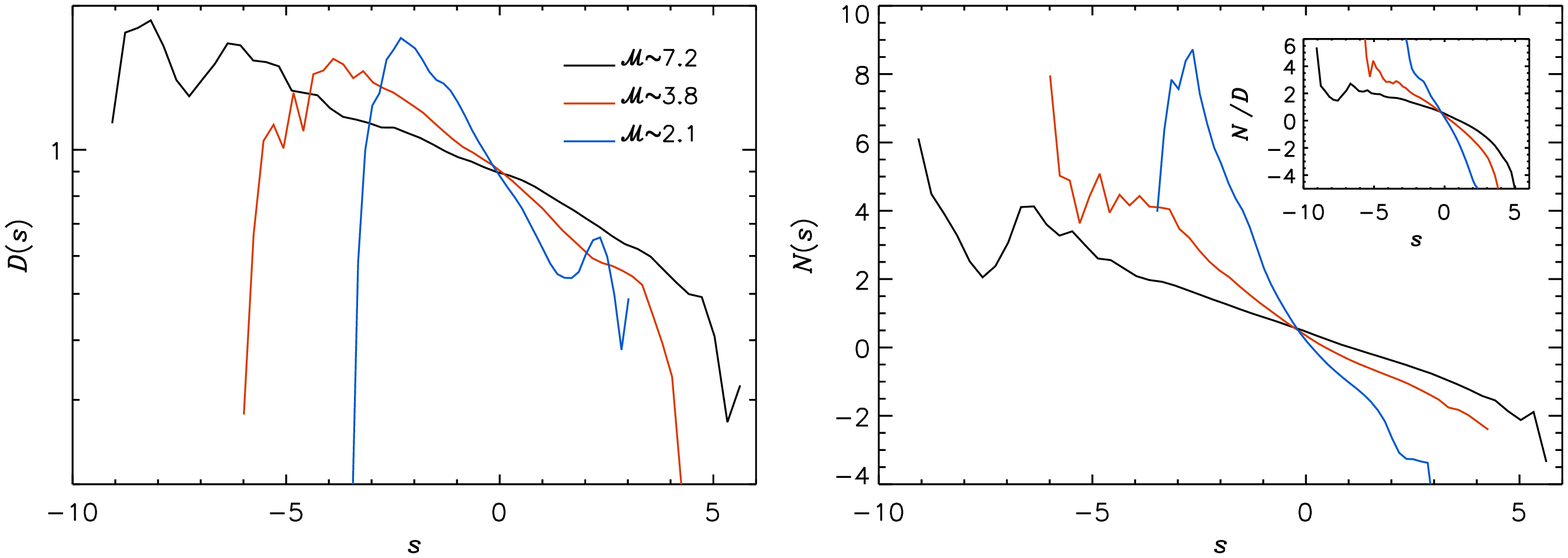}
\caption{ ${D}$ (left panel) and $ {N}$ (right panel) in 1024$^3$ N-S simulations at three Mach numbers. 
They were used in the calculation of the derived PDFs in the left panel of Figure 1. The inset shows 
the $N$ to $D$ ratio.}
\label{figpdf}
\end{figure*}

\subsection{The $D$ and $N$ terms}

The derived PDF from Eq.\ (\ref{solution}) is determined by the behaviors of $D$ and $N$ with $s$. In Fig.\ 2, we plot 
$D$ (left panel) and $N$ (right panel) measured from the N-S simulations, which were used to calculate the derived PDFs in the left panel of Figure 1. 
The left panel shows that $D$ decreases roughly exponentially with $s$. The oscillations and the fast drop of $D$ at 
extreme values of $s$ are likely due to the small sample size and insufficient statistics. 
The decrease of $D$ with $s$ may appear surprising, since the dense regions are 
usually associated with strong shocks where the divergence is large. 
In order to correctly understand the contribution of shocks to $D$, one needs to 
examine the internal structure of shocks, as the divergence $\nabla \cdot {\bs u}$ 
is small outside the shocks. Note that, since the density varies from a small value in the 
preshock region to a large postshock value, a strong shock contributes to $D$ for both large 
and small $s$.  On the other hand, due to the smaller density jumps, the weaker shocks 
with smaller divergence only contribute to $D$ at small densities. 
As the contribution to $D$ at large $s$ is from strong shocks, one might expect $D$ 
to be larger at higher density.  That, however, is not true, Within a shock structure, high density 
is located toward the postshock region where the divergence is small. This tends to make $D$ smaller 
at large density. It is the center of the shock that has the largest divergence, but the density there is 
relatively low. Furthermore, the highest density in the flow is likely in the postshock regions, where the 
divergence is small. This explains why $D$ decreases with $s$. 
 
With increasing Mach number, the decrease of $D$ with increasing $s$ becomes flatter. 
At larger $\mathcal{M}$, the divergence has stronger fluctuations, so that both 
$\langle (\nabla\cdot {\bs u})^2|s\rangle$ and $\langle (\nabla\cdot {\bs u})^2\rangle$ increase. 
However, the amplitude of their ratio, i.e., $D$, remains largely unchanged (as seen in the left panel). 
The ratio varies over a more entended range of $s$ because of the larger amplitude of density fluctuations, and thus the slope of $D$ becomes flatter with 
increasing $\mathcal{M}$. The flattening $D$ with $\mathcal{M}$ may be understood in an alternative way. 
From the definition of $D$, we have $\int_{-\infty}^{\infty} D(s) f(s) ds = \int_{-\infty}^{\infty} \langle (\nabla\cdot {\bs u})^2|s\rangle  
f(s) ds/ \langle (\nabla\cdot {\bs u})^2\rangle =1$. Therefore, as the PDF, $f(s)$, broadens with increasing Mach number,
 it is easy to show that, to satisfy the constraint $\int_{-\infty}^{\infty} D(s) f(s) ds=1$, $D(s)$ as a function of $s$ must become flatter.  
At $\mathcal{M} =7.2$, the decrease of $D$ with $s$ is already very slow, varying by a factor of a few over a 
broad density range. We expect that, as $\mathcal{M}$ keeps increasing, $D(s)$ would finally be close 
to 1 at all $s$, meaning that the amplitude of divergence is independent of the flow density in the large Mach number limit. 
 
The right panel of Figure 2 shows $N$ measured from the simulation data. 
The Lagrangian time derivative of the divergence in $N$ was computed 
as $d(\nabla \cdot {\bs u})/dt=  \nabla \cdot (\partial_t {\bs u}) + \bs u \cdot  \nabla (\nabla \cdot {\bs u})$.  
As seen in the right panel,  $N$ is positive at small densities but negative at large $s$. 
Considering that $N \propto -d(\nabla \cdot {\bs u})/dt$, a negative $N$ 
at large $s$ implies that the amplitude of the divergence of a flow element with large density 
tends to be reduced if the element is contracting with a negative divergence. 
The density of the contracting element increases, but the amplitude of its divergence would on average keep decreasing 
due to the more negative $N$ at larger $s$. With time, the divergence would eventually turn positive, and the density of 
the element will start decreasing, moving to smaller and smaller $s$. 
This process suggests that $N$ must be negative at larger $s$. Otherwise, a contracting fluid element at large $s$ would tend to 
experience stronger and stronger compressions, so that either the right PDF tail would be unphysically broad or the PDF would 
never reach steady state.  A similar argument applied to negative $s$ shows that $N$ must be positive at small densities.

With increasing Mach number, the  slope of $N$ becomes flatter and gets closer to zero. Based on the discussion above, this would allow a contracting 
(expanding) flow element to move to higher maximum density (or smaller minimum density),   
leading to a broader PDF at larger Mach number.   Note that if $N$ were 0, the PDF would be given by $\propto \exp(-s)/D(s)$, 
suggesting that the PDF tends to be negatively skewed in the large Mach number limit. 

The ${N}$ term enters our formula, Eq.\ (\ref{solution}), only through the ratio, $ {N/D}$, which is shown in the inset of 
the right panel.  An inspection of the factor $\exp(\int_0^s  {N(s')/D(s')} ds')$ in Eq.\ (\ref{solution}) reveals that 
an increase (or decrease) of $N/D$ at $s>0$ tends to increase (or decrease) the probability of the right wing 
of the PDF. On the other hand, an increase  (or decrease) of the ratio at $s<0$ would make the left wing narrower (or broader). 
The inset shows that, as the Mach number increases, the ratio ${N/D}$ decreases 
at $s<0$, but increases at $s>0$, and thus both the left and right  wings of the PDF 
broaden with increasing $\mathcal{M}$, as seen in Figure 1. In other words, the density PDF becomes broader, 
as the slope of $N/D$ as a function of $s$ becomes shallower with increasing $\mathcal{M}$. 

\begin{figure*}
\includegraphics[width=2\columnwidth]{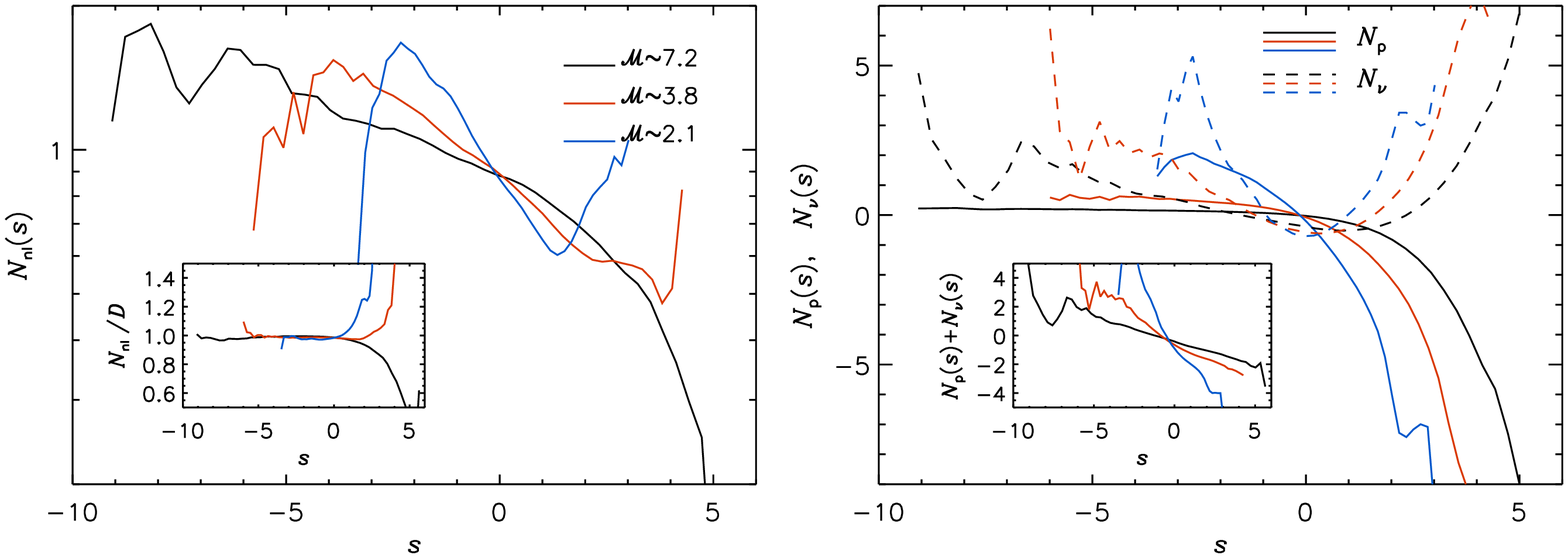}
\caption[]{Effects of dynamical processes on the density PDF in 1024$^3$ N-S simulations. 
As in Figures 1 and 2, blue, red and black lines correspond to $\mathcal{M}\simeq 2.1$ ,  3.8,  and 7.2, respectively. 
The left panel plots the nonlinear term, $N_{\rm nl}$,  whereas the right panel shows the pressure ($N_{\rm p}$, solid lines) 
and viscous ($N_{\nu}$, dashed lines) terms. The inset of the right panel gives the sum of pressure and viscous terms.}
\label{figpdf}
\end{figure*}

If the ratio $ {N/D}$ decreases linearly with $s$, then $\exp(\int_0^s  {N(s')/D(s')} ds')$ in Equation (\ref{solution}) is Gaussian, and its product with the exponential factor $\exp(-s)$ 
and $D^{-1}$, which is also roughly exponential (see the left panel), would result in a Gaussian 
distribution for $f(s)$. As seen in the inset, this is indeed the case for the Mach 2.1 flow, where the density PDF is close to lognormal. 
At larger Mach numbers, ${N/D}$ is not linear with $s$, and the density PDF is no longer well fit by a lognormal distribution. 
In particular, for $\mathcal{M}=3.8$ and $7.2$,  the ratio drops faster than linearly at $s>0$, especially toward the largest $s$, 
corresponding to a fast decrease of the probability at the right PDF tail.  The faster drop of the ratio at large
$s$ makes the PDF negatively skewed, as seen in Figure 1.  
We note that, at Mach 7.2, $N$ is almost linear with $s$ and it is the decrease of $D$ with $s$ (see the left panel) that causes 
the faster  drop of the ratio, $N/D$, at large $s$, leading to significant skewness of the PDF.  
N-S simulations at higher Mach numbers are needed to better see the trends of $D$ and $N$ with increasing $\mathcal{M}$, 
which would help us understand the PDF shape, in particular, the skewness as a function of $\mathcal{M}$.   

According to Equation (\ref{N-eq}), the behavior of $N$ is determined by various dynamical 
processes.  We next examine the effect of each process on the density PDF by analyzing its contribution to $N$.

\subsection{The dynamical effects}
 
As solenoidal driving is adopted, $N_{\rm a}=0$ in our simulations, and only the nonlinear,  
$N_{\rm nl}$, pressure, $N_{\rm p}$, and viscosity, $N_{\nu}$, terms contribute to $N$. An analysis 
of these three terms provides physical insights into the density PDF.  

In general, if a dynamical process gives a positive contribution to $N$ at $s>0$, it would amplify the right 
tail of the PDF at $s>0$, whereas a positive contribution at $s<0$ reduces the probability at the left tail. 
As mentioned earlier, this can be seen by inspecting the integral, $\int_0^s  {N(s')/D(s')} ds'$, in Equation (\ref{solution}). To understand this 
physically, we consider a flow region where a dynamical process provides a positive contribution to $N$. 
The region may be either expanding or contracting. In the probability space of $s$, an expanding or contracting region would move to lower or higher 
values of $s$, respectively. As $N$ is defined as the conditional average of $-d (\nabla \cdot {\bs u})/dt$, 
a positive contribution to $N$ corresponds to an increase of the divergence amplitude if the region is converging ($\nabla \cdot {\bs u} <0$),  
or a reduction of the divergence if the region is expanding ($\nabla \cdot {\bs u}>0$).  In other words,  
a positive contribution to $N$ indicates either an acceleration of contraction or a deceleration 
of expansion.  This means that in  the probability space, the dynamical process that provides a positive contribution to $N$ 
tends to increase the flux of probability to larger density or decrease the flux to lower density, causing the PDF to shift toward higher densities. 
Clearly, such a shift leads to the broadening of the right PDF tail at $s >0$, or the narrowing of the left tail at $s<0$.  
Similarly, it is expected that a negative contribution to $N$ reduces the right tail at $s>0$, but amplifies the left tail at $s<0$. 


We now consider the three dynamical processes one by one, starting with the nonlinear term, which is shown in the left panel of Fig.\ 3.  $ {N}_{\rm nl}$ is positive at 
all densities, and thus, as discussed above, the nonlinear process tends to shift the PDF toward higher density, 
making the right PDF tail broader and the left tail narrower. To understand why $ {N}_{\rm nl}$ is 
positive, we notice that for one-dimensional velocity structures like shocks, $ \partial_i u_j = \partial_j u_i$, 
so that ${N}_{\rm nl} \propto  \langle \partial_i u_j \partial_j u_i |s\rangle \ge 0$. Therefore, the nonlinear term would 
be positive if its main contribution were from shocks. The same argument applies to spherically symmetric flow structures, 
where $ \partial_i u_j$ is also equal to $\partial_j u_i$. The finding of ${N}_{\rm nl}>0$ suggests that the nonlinear term in the 
divergence equation tends to increase the amplitude of divergence in converging  regions, 
which is expected as the nonlinear effect is responsible for self-amplifying velocity gradients, 
or reduce the divergence amplitude in expanding regions.  

Except at the largest values of $s$ where the statistics is noisy, the behavior of $ {N}_{\rm nl}$ as a function of $s$ is 
very close to $ {D}$ for $s<0$. As seen in the inset, the ratio of $ {N}_{\rm nl}$ and  $D$ is close to 1. 
The near equality of $ {N}_{\rm nl}$ and  $D$ is easy to understand if the contributions to both $ {N}_{\rm nl}$ and 
$D$ are dominated by one-dimensional or spherically symmetric structures. As mentioned above, in such structures 
we have $\partial_i u_j \partial_j u_i  = (\partial_i u_i)^2$, so that 
${N}_{\rm nl} (s) \simeq  {D}(s)$.  Furthermore, assuming statistical homogeneity would give 
a constraint between ${N}_{\rm nl} (s)$ and  ${D} (s)$. It follows from statistical homogeneity 
that $\langle \partial_i u_j \partial_j u_i \rangle  = \langle (\partial_i u_i)^2 \rangle$, and rewriting it 
as $\int_{-\infty}^{+\infty}  {N}_{\rm nl} (s') f(s')ds' = \int_{-\infty}^{+\infty}  {D}_{\rm nl} (s') f(s')ds'$ shows that 
the averages of ${N}_{\rm nl} (s)$ and  ${D}_{\rm nl} (s)$ over the distribution of $s$ must be equal. 

The arguments above provide only a qualitative explanation for the near 
equality between $ {N}_{\rm nl}$  and $ {D}$, and it remains to be proved 
whether or not ${N}_{\rm nl} $ is exactly equal to ${D}$. If $ {N}_{\rm nl} =  {D}$ 
exactly holds, then our Equation (\ref{solution2}) becomes, 
\begin{equation}
f(s) = \frac{ C} {D(s)}\exp\left( \int\limits_0^{s}  \frac {N_{\rm p} (s') +N_{\nu} (s')}{D(s')} ds' \right). 
\label{solution3}
\end{equation}
Considering that the dependence of $D$ on $s$ is quite weak, especially at large Mach numbers, 
the equation suggests that the shape of the density PDF is determined primarily by the pressure 
and viscosity effects.  

The right panel shows the pressure (solid) and viscous (dashed) contributions to ${N}$. It turns out that  for all Mach numbers  
the pressure term, $N_{\rm p}$, is zero around $s=0$. $N_{\rm p}$ is positive and negative at  $s<0$ and $s>0$, respectively, 
meaning that it tends to make both the left and right PDF tails narrower.  
In a region of large density, the pressure is likely a local maxima, and $N_{\rm p}$ is expected 
to be negative at large $s$, since it is essentially defined as the conditional average of 
the Laplacian of $s$ for isothermal flows. A local pressure maxima would tend 
to expand the region and reduce the local density.  In other words, 
the pressure  suppresses strong density fluctuations. In particular, $N_{\rm p}$ drops rapidly at large positive 
$s$, and thus has a strong effect of making the right PDF tail 
narrower. On the other hand, the regions of low density are likely pressure valleys, corresponding to a 
positive $N_{\rm p}$ at negative $s$. The higher pressure from the surrounding gas would tend to increase the 
density,  causing a reduction of the left tail.  With increasing Mach number,  $N_{\rm p}$ gets closer to zero for 
both negative and positive $s$.  
Note that at larger $\mathcal{M}$, the pressure, $p$, is actually stronger, but the amplitude of $N_{\rm p}$, which 
is defined as a ratio $C_{\rm s}^2 \langle \nabla^2 s|s \rangle/ \langle (\nabla \cdot {\bs u})^2\rangle$, is smaller 
due to the stronger fluctuations of divergence. 
The weaker effect of $N_{\rm p}$ at larger Mach number would lead to a broader PDF. For $\mathcal{M} = 3.8$ and 7.2, $N_{\rm p}$ is only 
slightly positive at $s<0$, suggesting that the pressure plays a minor or negligible role on the left PDF tail at high Mach numbers.    


The dashed lines in the right panel of Figure 3 plot the viscous term ${N}_{\nu}$, which shows a quite 
complicated behavior. ${N}_{\nu}$ is positive at both large and small $s$, but negative at intermediate densities around 
$s = 0$.  
As a function of $s$, ${N}_{\nu}$ must switch sign due to a constraint, $\int_{-\infty}^{+\infty} {N}_{\nu} (s) f(s) ds=0$, which follows from statistical homogeneity\footnote{Similarly, the 
constraint also applies to the pressure term, i.e., $\int_{-\infty}^{+\infty} {N}_{\rm p} (s) f(s) ds=0$.}.  By definition, $\int_{-\infty}^{+\infty} {N}_{\nu} (s) f(s) ds \propto -\langle \nabla \cdot {\bs a}_{\nu} \rangle$, 
where ${\bs a}_{\nu} = \rho^{-1} \nabla \cdot \sigma$ is the viscous term in the N-S equation (\ref{v-eq}). Clearly,  statistical homogeneity requires $\langle \nabla \cdot {\bs a}_{\nu} \rangle =0 $, 
so that $\int_{-\infty}^{+\infty} {N}_{\nu} (s) f(s) ds$ must vanish. 

To better understand the viscous term,  we separate the contributions to ${N}_{\nu}$ at a given $s$ from regions with positive and 
negative divergence. The two contributions, denoted as ${N}_{\nu}^+$ and  ${N}_{\nu}^-$, are defined such that ${N}_{\nu} = {N}_{\nu}^+ +{N}_{\nu}^-$. 
As shown in Figure 4, ${N}_{\nu}^+ $ from the expanding regions is positive for all $s$. This can be intuitively understood as the viscosity slowing down 
the expansion of regions with positive divergence. 
On the other hand, the behavior of ${N}_{\nu}^-$ is more complicated. In particular, it is surprising that ${N}_{\nu}^-$ is  
positive at large densities as well. In a dense region with negative divergence, it would be natural to expect that the viscosity slows down the contraction, 
and thus suppresses strong density fluctuations at large density. This is in contrast to the positive ${N}_{\nu}^-$ at large $s$, which tends to the 
amplify the right PDF tail. 

\begin{figure}
\includegraphics[width=1\columnwidth]{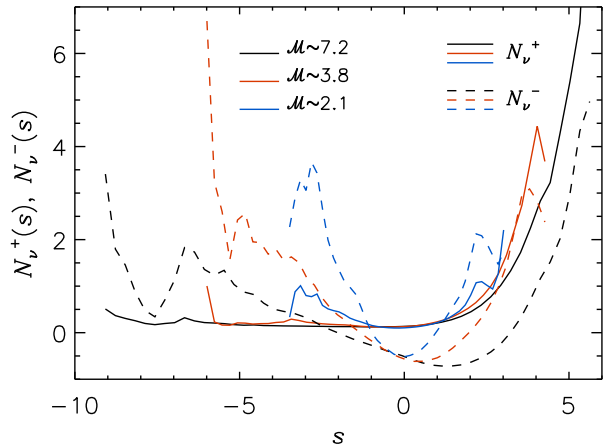}
\caption[]{Contributions to the viscous term, $N_{\nu}$, from  flow regions with positive (solid lines, $N_{\nu}^+$ ) and negative (dashed lines, $N_{\nu}^-$) divergence. 
Blue, red and black lines correspond to simulated flows with $\mathcal{M} =2.1$, 3.8 and 7.2, respectively.}
\label{figpdf}
\end{figure}

This complicated and counter-intuitive behavior of ${N}_{\nu}^-$ may be understood from its dependence on the third order derivatives of 
the velocity field. 
For illustration, we analyze the third order velocity derivatives inside shocks. 
Consider, for example, a static 1D shock, whose center is located at the origin.   
The velocity is assumed to lie in the $+x$ direction, decreasing from the preshock value at $x<0$ to the postshock value at $x>0$.   
We examine the behavior of the viscous term, $a_\nu(x)$, in the N-S equation across the shock. 
For simplicity, we assume that $a_\nu(x)$ is given by the second-order derivative of the velocity field\footnote{With the viscous tensor adopted in our simulations, $a_\nu$ depends not only on the 
velocity derivative but also on the density derivative. However, the behavior of $a_\nu$ described here based on the analysis of the second-order velocity derivative is 
qualitatively consistent with a numerical solution for a 1D static shock using the same viscous term as adopted in the simulations.}, i.e., $a_\nu(x) \propto \partial^2 u(x)/\partial x^2$. 
Due to the second-order derivative, $a_\nu(x)$ is negative from the preshock region to the shock center because the velocity profile in this range is concave. 
On the other hand, from the shock center to the postshock region $a_\nu(x)$ is 
positive due to the convex velocity profile, meaning that the viscosity accelerates rather than decelerates 
the velocity. This may appear to be strange because overall the velocity is decelerating. 
The deceleration of the velocity in this region is actually due to the pressure term. 
The viscous acceleration, $a_\nu$, is expected to switch sign, when the velocity profile changes from concave to convex somewhere 
around the shock center. To summarize, $a_\nu$ first decreases from 0 in the preshock region to a minimum value at some point, $x_{\rm min}$, 
then it starts to increase, passes 0 and rises to a maximum value, say at $x_{\rm max}$, before finally dropping to 0 in the 
postshock region.
 
We then take the spatial derivative or divergence of $a_\nu(x)$. Clearly, at $x \le x_{\rm min}$, the derivative of $a_\nu(x)$ with 
respect to $x$ is negative, and thus the contribution from this low-density shoulder to $N_{\nu}$ is positive since 
$N_{\nu} \propto - \partial_x a_{\nu} (x)$ by definition. Physically, in this region the viscosity ``accelerates" the velocity 
deceleration, and thus tends to increase the density. In the central shock region from $x_{\rm min}$ to $x_{\rm max}$, $a_{\nu} (x)$ 
increases with $x$, so that $\partial_x a_{\nu} (x)>0$. The contribution of the central region to $N_{\nu}$ is thus 
negative, consistent with negative $N_{\nu}^-$ around $s\simeq 0$ as observed in Figure 4. 
Finally, in the high-density shoulder with $x>x_{\rm max}$,  $a_{\nu}$ 
decreases from a positive maximum value to 0 in the postshock region, so that $\partial_x a_{\nu} (x)<0$. 
A negative divergence of $a_{\nu}$ in the high-density shoulder of the shock tends to increase the amplitude of 
the velocity divergence there and gives a positive contribution to $N_{\nu}^-$. This provides an explanation to the counter-intuitive 
result that $N_{\nu}^-$ is positive at large $s$, as seen in Figure 4.  
Since both $N_{\nu}^-$ and $N_{\nu}^+$ are positive at large densities,  
$N_{\nu}$ is positive at large $s$ and tends to amplify the right PDF tail. 

The inset in the right panel of Figure 3 shows the sum, $N_{\rm p} + N_{\nu}$, of the pressure  and viscous terms. 
As discussed earlier, if $N_{\rm nl} = D$, it is the sum, $N_{\rm p} + N_{\nu}$,  and $D$ that determine the PDF (see Equation \ref{solution3}).  
At large $s$, the effect of pressure to suppress strong density fluctuations dominates over the effect of $N_\nu$ to amplify 
the right tail, and the sum is negative.  At low densities, both $N_{\rm p}$ and $N_{\nu}$ contribute to reduce the left tail, 
and we see that at large $\mathcal{M}$ the effect of $N_{\nu}$ at small $s$ is stronger than $N_{\rm p}$. 

In summary, using numerical simulations that explicitly include physical viscosity in combination with our theoretical formulation 
for the density PDF, we found that the nonlinear effect tends to shift the PDF toward large densities, 
amplifying the right PDF tail and reducing the left tail. The pressure term suppresses both the left and right tails, although its effect 
on the left tail is minor for large Mach numbers.  The pressure is the only process that reduces the right tail. 
Surprisingly, the viscous term tends to amplify the right PDF tail.  This counter intuitive effect is 
caused by the complicated behavior of the third order spatial derivative in $N_\nu$. As an example,  
the viscosity tends to increase the amplitude of the velocity divergence in the high-density shoulders of shocks.   
The analysis of the effects of various dynamical processes based on our theoretical formulation reveals 
that the origin of the density PDF in supersonic turbulence is very complicated, suggesting that 
simple phenomenological models may not be able to fully capture the shape of the PDF with high accuracy. 
 
\section{The accuracy of density PDF in the Riemann solver simulations} 

Most simulations of supersonic interstellar turbulence are based on Riemann solvers that evolve the Euler equation. 
Without an explicit physical viscosity\footnote{The validity of neglecting the physical viscous term relies on the 
assumption that the inertial-range statistics of turbulence are universal and independent of the exact form of viscosity 
that operates at the smallest scales.}, the Riemann simulations may acquire the largest effective Reynolds number 
possible at a given numerical resolution, resulting in a more extended inertial range and better resolved high-density fluctuations 
(at large Mach numbers) than in N-S simulations of the same size.

\begin{figure}
\includegraphics[width=1\columnwidth]{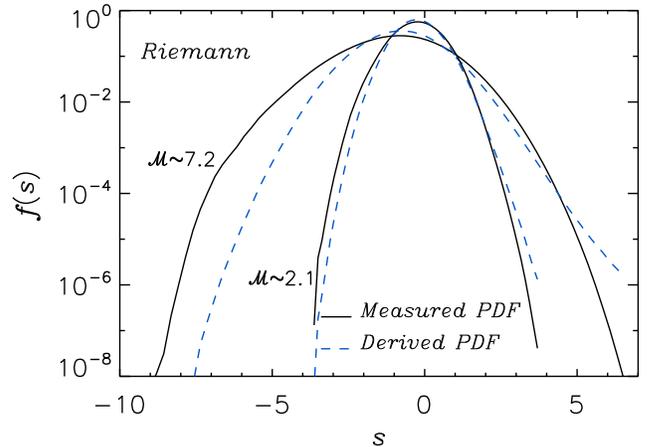}
\caption[]{Density PDFs at  $\mathcal{M} =2.1$ and 7.2 from 1024$^3$ Riemann runs. 
The solid and dashed lines show the directly measured PDFs and  the derived PDFs from  
Equation (\ref{solution}), respectively.  The derived PDFs are calculated with the velocity 
divergence computed from the simulation data, and the discrepancy between solid and dashed 
lines indicates inaccuracy of the divergence in Riemann runs.}
\label{pdf: Riemann}
\end{figure}

However, Riemann-solver solutions of supersonic flows contain near-discontinuities.
As a result, spatial derivatives computed from the solution cell-centered values
are inaccurate, resulting in an apparent violation of the continuity equation (even if the numerical scheme enforces
the conservation of mass), as recently shown by Pan et al. (2019). 
A stark example of such inaccuracy is illustrated 
by Figure \ref{pdf: Riemann}, where we show the measured PDFs (solid lines)  in two 1024$^3$ Riemann simulations 
with $\mathcal{M} =2.1$ and 7.2. The figure also shows, as short-dashed lines, the derived
PDFs from Equation (\ref{solution}), derived in \S 2.1. While the derived PDF matches nearly exactly the PDF
measured in the N-S simulation (left panel), as already shown in the left panel of Figure \ref{pdfs}, for the Riemann run
(right panel), the PDF derived from Equation (\ref{solution}) deviates strongly from the measured PDF. This strong discrepancy
is a consequence of the inaccuracy of the velocity divergence measured from the Riemann simulations (Pan et al. 2019).

\begin{figure*}
\includegraphics[width=2\columnwidth]{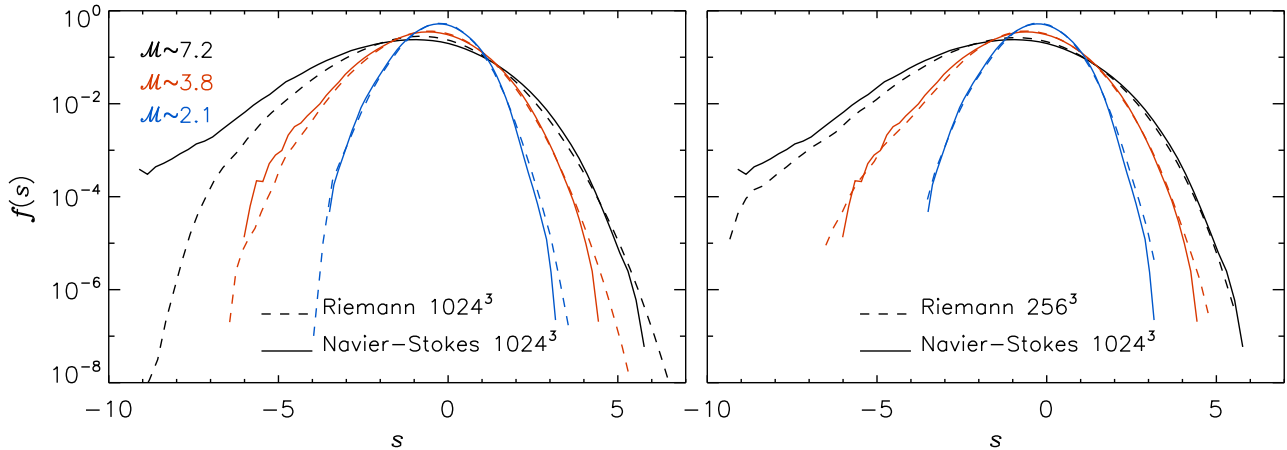}
\caption[]{Measured density PDFs in N-S (solid) and  Riemann
 (dashed) simulations at three Mach numbers, $\mathcal{M} = 2.1$ (blue), 3.8 (red) 
and 7.2 (black). Solid lines in both panels correspond to the N-S runs at 
1024$^3$. Dashed lines in the left and right panels show the PDFs from Riemann runs at 1024$^3$ and $256^3$, respectively.}
\label{pdf:RiemmanVsNS}
\end{figure*}


The inaccuracy of the spatial derivatives and the apparent violation of the continuity equation 
raise the issue of the reliability of the density PDFs computed from the Riemann simulations. To address this issue, we compare these PDFs 
with those from the N-S runs at the same Mach numbers. Figure \ref{pdf:RiemmanVsNS} shows the PDFs measured in 
the N-S (solid) and Riemann (dashed) simulations at three Mach numbers, 2.1,  3.8 and 7.2. 
In the left panel, the solid and dashed lines, corresponding to the N-S and Riemann runs at the same resolution of 
1024$^3$, are in good general agreement. In particular, the PDFs at Mach 2.1 match quite well. 
The far right tail of the PDF from the Riemann runs is always higher than that from N-S runs 
probably due to the higher effective numerical resolution. At Mach 3.8 and 7.2, the left PDF tail from the 
N-S runs is broader than in the Riemann runs. The left tail of the density PDF is more 
uncertain than the right tail, as it is more sensitive to the pattern of the driving force 
and shows stronger variations with time\footnote{It is well known that the statistical relaxation of the left tail is hard
 to achieve, requiring long time averaging to reach convergence. With extensive experimentation, we found that, 
 even for solenoidal driving, different choices for 
the correlation time and probability distribution of the driving term lead to differences in the left tail. 
The reason might be that regions of low density are likely to have large sizes and thus may be more sensitive to the direct effect of the large-scale driving force.}. 
 Although far from being perfect, 
the agreement with the results of the N-S runs provides evidence for the reliability of 
the density PDF from the Riemann simulations.    

Due to the explicit viscosity, the Reynolds number in an N-S run is significantly smaller than the 
effective value in the corresponding Riemann run at the same resolution. As seen in Figure 1 of Pan et al.\  (2019), 
the simulated flow from an N-S run is much smoother than that from a Riemann run.
Therefore, it is more appropriate to compare the N-S runs with the Riemann simulations at a lower numerical 
resolution. The right panel of Figure \ref{pdf:RiemmanVsNS} shows the PDFs measured from the Riemann runs at $256^3$ are 
in good agreement with the results of the N-S simulations at $1024^3$. 
In comparison to the left panel, the agreement is significantly improved\footnote{A comparison of the dashed lines in the two panels reveals that the PDF tends to become more symmetrical as the resolution increases from 256$^3$ to 1024$^3$.}, further confirming the 
reliability of the PDFs measured from the Riemann simulations.  It also shows 
that, by neglecting the viscous term, the Riemann-solver simulations achieve an effective spatial 
resolution approximately four times larger than that of the N-S simulations.   

One might suspect that the inaccuracy of density gradients in the Riemann simulations could affect the density PDF to 
some degree, but the agreement between the PDFs from the N-S and the Riemann simulations indicates that this is not a significant
problem.

Further insight into the accuracy of the density PDFs from the Riemann runs may be provided 
by 
the PDF formula of \citet{Pope+Ching93}. 
A derivation of the Pope-Ching formula, 
Equation (\ref{stimesolution2-eq}), is given in Appendix A, and its validity 
is verified for both the N-S and Riemann simulations in Appendix C.  
The formula depends only on the conditional expectations of the first and second order time derivatives of $s$, suggesting that the PDF 
of the density may be determined by how it evolves with time. If in two simulated flows the evolution of $s$ 
at each given density level is statistically the same, then the density PDFs in the two flows would be the same 
(see Equation (\ref{stimesolution2-eq}) ). In Appendix D, we compare $\langle \dot{s}|s\rangle$ and $\langle \ddot{s}|s\rangle$ 
from the N-S and Riemann runs at Mach 7.2 and show that their dependence on $s$ is
statistically similar in the two runs (see Figure \ref{pdf: timederivatives}). This means that, 
despite the numerical artifacts in the spatial derivatives due to the absence of a physical viscosity, the evolution of the density in the 
Riemann runs is a good approximation to what happens in a real flow, which would explain why the density PDFs in the N-S and 
Riemann runs are similar, as seen Figure \ref{pdf:RiemmanVsNS}. 

To summarize this section, we have shown that the PDFs measured from the Riemann runs 
are reliable, even though the code does not explicitly include a physical viscous term. In what follows, 
we will use the Riemann runs to examine the Mach dependence of the density PDF over a range of Mach 
number values that would not be accessible with the lower effective resolution of the N-S simulations.
 
\section{The Mach number dependence of the density PDF}

\begin{figure}
\includegraphics[width=1\columnwidth]{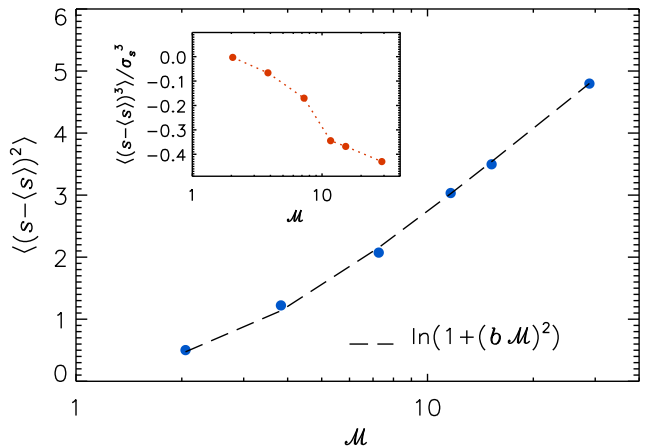}
\caption[]{Variance of the PDF of $s$ from Riemann simulations at different 
Mach numbers. Dashed line plots the commonly-used 
variance-Mach number relation, $\sigma_{s}^2 = \ln (1+b^2 {\mathcal M}^2)$, with $b = 0.38$. The inset shows 
the skewness of the PDF of $s$ as a function of Mach number.  
}
\label{pdf: varianceMach.eps}
\end{figure}

\begin{figure*}
\includegraphics[width=2\columnwidth]{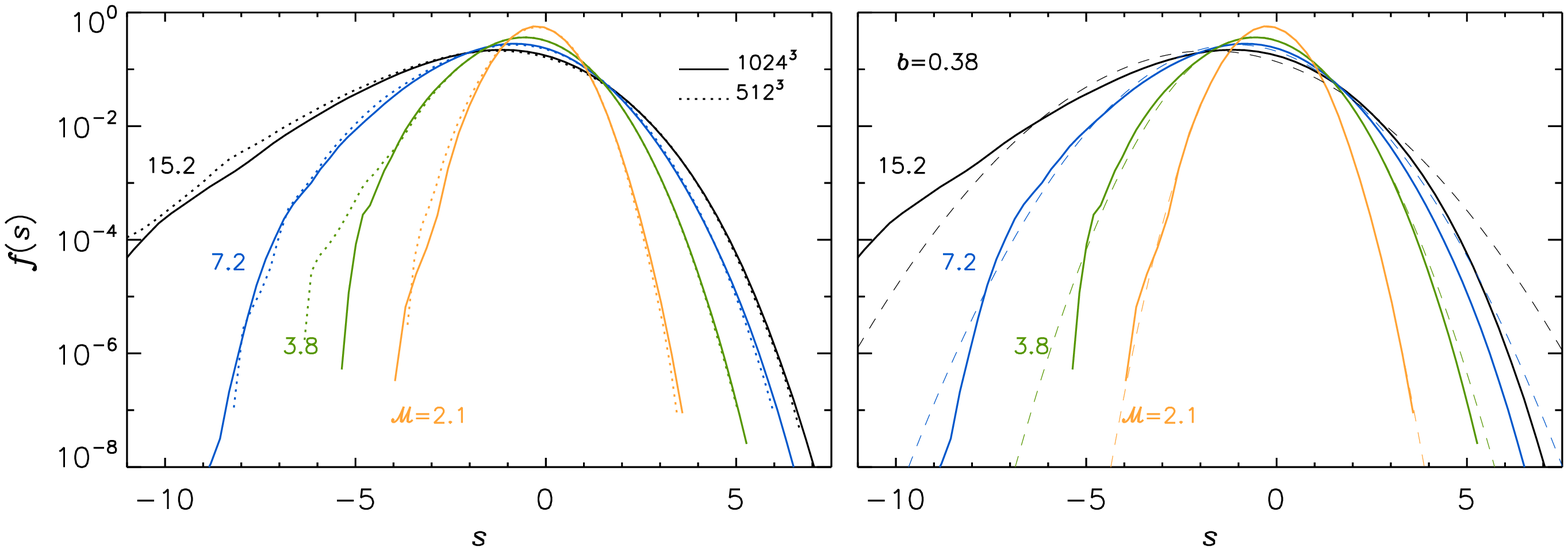}
\caption[]{Density PDFs measured from Riemann runs at four approximately logarithmically-spaced Mach number values.  
In both panels, the solid lines correspond to the PDFs from $1024^3$ runs. 
Dotted lines in the left panel plot the PDFs from $512^3$ runs, whereas dashed lines in the right panel 
are lognormal distributions with $\sigma_{s}^2 = \ln (1+b^2 {\mathcal M}^2)$, using the best-fit value $b=0.38$.  
}
\label{PDFRiemann.eps}
\end{figure*}

The dependence of the density PDF on the Mach number is of great astrophysical interest, as the 
density PDF is an essential element in theoretical models of star formation based on turbulent fragmentation 
(see references in \S1). Galactic molecular clouds (MCs) are characterized by turbulent motions with Mach numbers 
(defined with the three-dimensional rms velocity, as in the simulations), in the approximate range of ${\mathcal M}\approx 20-80$,
based on well-known velocity-size relations \citep[e.g.][]{Larson81,Solomon+87,Heyer+Brunt04}. More extreme
star-formation environments such as starbursts and the progenitors of early type galaxies may be characterized 
by even larger Mach numbers \citep[e.g.][]{Hopkins13,Chabrier+14}. However, most existing parameter studies of statistical 
properties of supersonic turbulence are based on unigrid simulations with a maximum resolution typically not exceeding $1024^3$
computational elements, where the high-density tail of the density PDF cannot be numerically converged for ${\mathcal M}\gsim 10-20$.

One way to obtain the density PDF at realistically high Mach numbers, such as in the largest MCs or in extreme star-formation
environments, is by extrapolation based on the trend of the PDF at smaller ${\mathcal M}$. In star formation models, such extrapolation 
assumes that the density PDF is lognormal with its width fixed by a variance-Mach number relation. In this section, we will test 
the accuracy of this commonly-adopted prescription against the results of Riemann simulations with ${\mathcal M}$ up to $\approx 30$.  

Extensive numerical studies of isothermal supersonic turbulence have established a relation 
between the rms width, $\sigma_{s}$, of the PDF of $s$ and the Mach number,
\begin{equation}
\sigma_{s}^2 = \ln (1+b^2 {\mathcal M}^2), 
\label{eq:varianceMach}
\end{equation}
where $b$ is a fitting parameter \citep{Padoan+97imf}. 
The best-fit parameter, $b$, varies from study to study, depending both on the adopted numerical code and the simulation setup, 
and on a number of physical factors, such as the driving pattern \citep{Federrath+10}, the presence of magnetic fields \citep{Molina+12}, 
and the adopted equation of state \citep{Nolan+15}.  

Figure \ref{pdf: varianceMach.eps} shows that Equation (\ref{eq:varianceMach}) with a fixed value of $b$ matches well the results
of our Riemann simulations at six Mach number values, 2.1, 3.8, 7.2, 11.6, 15.2 and 29.8.  
The dashed line plots Equation (\ref{eq:varianceMach}) with $b = 0.38$. Consistent with 
previous studies, Equation (\ref{eq:varianceMach}) provides a good fit to the 
variance-Mach number relation, and the best-fit value, $b = 0.38$, is in general agreement with the value of $\sim 1/3$ 
found in simulations with solenoidal driving \citep[e.g.][]{Federrath+08,Price+11}. The inset plots the skewness of 
the PDF as a function of $\mathcal{M}$. At small $\mathcal{M}=2.1$, the PDF of $s$ is approximately 
Gaussian, so the skewness is close to zero.      
At larger Mach numbers, the PDF is no longer symmetric and becomes negatively skewed.   
As seen in the inset, the skewness gets stronger with increasing Mach number. 
The skewness of the density PDF in the case of solenoidal driving at high Mach number had been noticed before \citep[e.g.][]{Federrath13},
but its dependence on Mach number had not been studied so far.  


The left panel of Figure \ref{PDFRiemann.eps} shows the density PDFs 
measured from the Riemann runs at four Mach number values with approximate 
logarithmic spacing, 2.1, 3.8, 7.2 and 15.2. 
By comparing the PDFs at numerical resolutions of 512$^3$ and 1024$^3$, 
we see that numerical convergence is achieved except at far left tails. 
As mentioned earlier, the left PDF tail typically exhibits stronger 
temporal variations, and is harder to  relax statistically, perhaps 
because the low-density regions occupy large volumes and 
are thus more sensitive to the large-scale driving force. 

With increasing $\mathcal{M}$, the PDF becomes more and more 
skewed.  Interestingly, we find that the right PDF tail appears to be converging with increasing Mach number. 
Although the rms width of the PDF increases with  $\mathcal{M}$ (see Figure \ref{pdf: varianceMach.eps}), 
the negative skewness makes the broadening of the right PDF tail with increasing $\mathcal{M}$ quite 
slow. Based on the trend seen in Figure \ref{PDFRiemann.eps}, we 
would expect the right PDF tail to converge at sufficiently high Mach number, 
a result that would significantly impact the predictions of star-formation models 
based on turbulent fragmentation.  

The PDF of our Riemann simulation at $\mathcal{M}\approx 30$ confirms the tendency of 
the high-density tail to saturate towards higher Mach numbers. However, we have not plotted that PDF at $\mathcal{M} = 29.8$ in
Figure \ref{PDFRiemann.eps} because the comparison with the corresponding 512$^3$ simulation
shows that the right PDF tail is not fully converged at the 1024$^3$ resolution. Plotting the PDF
of the Mach 30 simulation in Figure \ref{PDFRiemann.eps} would give the misleading impression 
that the convergence of the right PDF tail with increasing Mach number is faster than it really is. 
Despite the lack of convergence of the high-density tail, the variance of the PDF at $\mathcal{M} \approx 30$ 
follows the general relation (\ref{eq:varianceMach}), as shown in Figure \ref{pdf: varianceMach.eps}, 
because the increase of the variance with Mach number is dominated by the central part of the PDF, rather than the far tails. 
Simulations at higher resolutions or with adaptive-mesh-refinement (AMR) methods will be pursued in a future work to
study the high-density PDF tail at Mach number $\gsim 30$.   

The dashed lines in the right panel of Figure \ref{PDFRiemann.eps} show 
lognormal distributions with $\sigma_{\rm s}$ given by Equation (\ref{eq:varianceMach}) with $b=0.38$.  
Combing the lognormal shape with a variance-Mach number relation is the common practice used in star-formation models to 
prescribe the density PDF in MCs.  
However, this commonly-used prescription fails to fit the measured PDFs in our 
Riemann simulations (solid lines), even though the variance-Mach number relation, Equation (\ref{eq:varianceMach}), is in 
good agreement with the simulation data (see Figure \ref{pdf: varianceMach.eps}).  
The main reason for the failure is that, at large Mach numbers, the measured PDF is 
highly skewed. Due to the negative 
skewness, the contribution to the variance of $s$ is mainly from the left wing of the PDF, 
and thus, with the variance alone, one cannot reproduce correctly the right PDF 
tail. The right panel of Figure \ref{PDFRiemann.eps} shows that the right tail of the PDF 
from the commonly-used prescription broadens much faster than the PDFs from our simulations,
and thus greatly overestimates the probability of finding regions of very high density. 
In contrast to the PDFs from our simulations, the dashed lines from the lognormal prescription 
show no sign of convergence with increasing Mach number. 
If the $1024^3$ results for $\mathcal{M} = 15.2$ are nearly converged, the standard lognormal model may overestimate the probability of gas density $\rho/\bar{\rho}\sim 200$
by more than a factor of 10, and the probability of gas density $\rho/\bar{\rho}\sim 700$ by more than a factor of 100. At larger
Mach numbers of $40-80$, typical of the largest MCs or extreme star-formation environments, it would overestimate
the density PDFs by several orders of magnitude at high densities.   

A fitting function for the density PDF measured in our simulations as a function of the Mach number 
would be of practical use for star formation models. We postpone the effort of obtaining such fitting functions to a 
future work, where we will carry out AMR simulations for both HD and MHD cases. 
AMR simulations can better resolve the right PDF tail, and make it possible to extend our study to 
higher Mach numbers, $\mathcal{M} \gsim 30$.  
With better resolution and 
larger Mach numbers, we expect the AMR simulations to provide a definite answer concerning the saturation 
of the right PDF tails with increasing $\mathcal{M}$.    

In summary, using simulations with Mach number up to 30, we confirmed the validity of the variance-Mach number relation, Equation (\ref{eq:varianceMach}), 
in isothermal turbulence. However, the commonly adopted prescription for the density PDF assuming a lognormal shape and using  
Equation (\ref{eq:varianceMach}) for the width fails to fit the simulation results and greatly overestimates the probability at the far right tail. 
The measured PDF from our simulations shows strong negative skewness at high Mach numbers, and its right tail appears to be converging with increasing Mach number.    

\section{Conclusions}

We have studied the PDF of density fluctuations in supersonic turbulence using 
both theoretical methods and numerical simulations. Using the probabilistic approach to 
turbulence, we have developed a theoretical formulation for the PDF of density fluctuations at steady state. 
We have carried out two sets of numerical simulations, one based on a finite-volume method to solve the Euler 
equation and the other based on a finite-difference method to solve the Navier-Stokes equation. 
A combination of the theoretical formulation and the simulation data has provided physical insights 
into the density PDF in supersonic turbulence.  We summarize the main conclusions as follows:  

\begin{enumerate}

\item  A theoretical formulation for the density PDF was established from first 
principles.  By deriving an exact equation for the density PDF from the continuity equation, 
we obtained two formulas, Equations (\ref{solution}) and (\ref{solution2}), for the density PDF at steady state, 
connecting the PDF to the conditional statistics of the flow divergence, and providing physical 
insights into the effects of the dynamical processes on the PDF shape.  The validity of our theoretical 
formulation was confirmed by our simulation data from  the N-S runs. 

\item  Combining the theoretical formulation and the simulation data from the N-S runs, 
we examined the effects of various dynamical processes on the PDF shape.  
We showed that the nonlinear term in the divergence equation tends to shift the density PDF toward larger densities, 
amplifying the right tail while reducing  the left one.  The pressure term has the effect of narrowing both the right and left 
tails. Counter-intuitively, we found that the effect of viscosity tends to broaden the right tail of the PDF.  

\item  Our study provided further evidence for a previous finding that the velocity divergence may not be 
reliably evaluated in simulations using Riemann solvers. In the Riemann simulations, the derived PDF from 
our formula, Equation (\ref{solution}), with the velocity divergence computed from the data, 
shows strong discrepancy with the directly measured PDF, indicating the inaccuracy of the velocity derivatives, as found in Pan et al.\ (2019). 

\item  By comparing against the results from the N-S runs,  we examined the reliability 
of the density PDF in Riemann simulations. We showed that, despite the numerical 
artifacts at the smallest scales, and in particular the apparent violation of the continuity equation, 
the PDF of density fluctuations in the Riemann runs is consistent with that 
measured from N-S runs. Furthermore, by neglecting the viscous term, the Riemann simulations achieve an effective spatial resolution 
approximately four times higher than the N-S runs of the same size. 

\item  Using Riemann simulations with Mach number up to $\simeq 30$, we investigated 
the dependence of the density PDF on ${\mathcal M}$. The width of the PDF is well fit by 
the variance-Mach number relation $\sigma_{s}^2 = \ln (1+b^2 {\mathcal M}^2)$ with $b= 0.38$. However, 
the commonly adopted prescription for the density PDF that combines a lognormal shape and 
the variance-Mach number relation for the PDF width fails to fit the simulation results, and in particular it greatly 
overestimates the probability at the right tail at large Mach numbers.  The PDF shows negative skewness that increases with increasing
${\mathcal M}$, and the growth of the right PDF tail with increasing ${\mathcal M}$ tends to saturate. We expect the right tail to 
converge to some asymptotic function at sufficiently high ${\mathcal M}$, which, if confirmed, would have a significant impact on 
theoretical models of star formation. 

\end{enumerate}

In order to provide accurate estimates of the density PDF at realistic Mach 
numbers in MCs, future studies are being planned to better understand the behavior 
of the density with increasing Mach number. Such studies will require AMR simulations to fully resolve the right PDF tail at $\mathcal{M} \gsim 30$, 
which would allow us to investigate the possible saturation of the right tail of the density PDF
towards an asymptotic solution at large Mach numbers. The determination
of such an asymptotic PDF would help better constrain the predictions of theoretical 
models of star formation.

\acknowledgements
LP acknowledges support from the Youth Program of the Thousand Talents Plan in China.
PP acknowledges support  by the Spanish MINECO under project AYA2017-88754-P. 
The work of {\AA}N was supported by grant 1323-00199B from the Danish Council for Independent 
Research (DFF). The Centre for Star and Planet Formation is funded by the Danish National Research Foundation (DNRF97).
Storage and computing resources at the University of Copenhagen
HPC centre, funded in part by Villum Fonden (VKR023406), were used to carry out the 
simulations. We thankfully acknowledge the computer resources at MareNostrum and the technical support provided by 
the Barcelona Supercomputing Center (AECT-2018-3-0019).

\appendix

\section{A. The Pope \& Ching (1993) Formulation} 

Pope \& Ching (1993)  derived a general formula for the PDF of any quantity in a statistically stationary process in terms of the conditional means 
of the time derivatives of the quantity. The formula can be applied to study the density fluctuations in supersonic turbulence at steady state (Nordlund \& Padoan 1999). 
The derivation of the formula starts with taking the first and second time-derivatives of the fine-grained PDF,  i.e., 
$\partial_t g(\zeta; {\bs x}, t)  =  -  {\partial_\zeta ( g(\zeta; {\bs x}, t)  \dot{s}) }$, 
and $\partial_t^2 g(\zeta; {\bs x}, t)  =  \partial_\zeta^2  [ g(\zeta; {\bs x}, t ) \dot{s}^2 ] -   \partial_\zeta 
[g(\zeta; {\bs x}, t) \ddot{s}] $, 
where $\dot{s} \equiv  {\partial_t s}$ and $\ddot{s} \equiv  {\partial_t^2 s} $, and we have used the fact that $\dot{s}$ and $\ddot{s}$ are independent of $\zeta$. 
Taking the average of the equation for the second order derivative and assuming statistical stationarity yields,
\begin{equation}
 \frac {\partial^2 \left[ f(\zeta)  \langle  \dot{s}^2|s=\zeta \rangle \right]}{\partial \zeta^2} =  \frac { \partial \left[ f(\zeta) \langle \ddot{ s} |s=\zeta \rangle\right]}{\partial \zeta},
\label{stimeave-eq}
\end{equation}
where $f(\zeta)$ is the coarse grained PDF at steady state. The equation is solved by, 
\begin{equation}
f(\zeta) = \frac{C}{\langle  \dot{s}^2|s=\zeta \rangle } \exp\left(\int\limits_{0}^{\zeta} \frac{\langle \ddot{ s} |s=\zeta' \rangle} {\langle  \dot{s}^2|s=\zeta' \rangle} d\zeta'  \right)
\label{stimesolution2-eq}
\end{equation}
with $C_{\rm pc}$ an integration constant fixed by the normalization of the PDF. We may rewrite the formula in a form similar to Equation (\ref{solution}) in \S 2, 
\begin{equation}
f(s) = \frac{C'}{D_{\rm PC} (s)} \exp\left(\int\limits_{0}^{s} \frac{N_{\rm PC}(s')} {D_{\rm PC} (s')} ds' \right).
\label{DNstimesolution-eq}
\end{equation}
where $N_{\rm PC}(s) = \langle \ddot{s} |s \rangle/\langle \dot{s}^2 \rangle $, and $D_{\rm PC} (s) = \langle \dot{s}^2 |s \rangle/\langle \dot{s}^2 \rangle $, and $C' =  C/\langle \dot{s}^2 \rangle$.
With the first and second time-derivatives of $s$ computed from the simulation data, one may derive the PDF of $s$. 
A comparison of the derived PDF using equation (\ref{DNstimesolution-eq}) with the directly measured PDF was presented in \S 5.

%

\begin{figure}
\includegraphics[width=1\columnwidth]{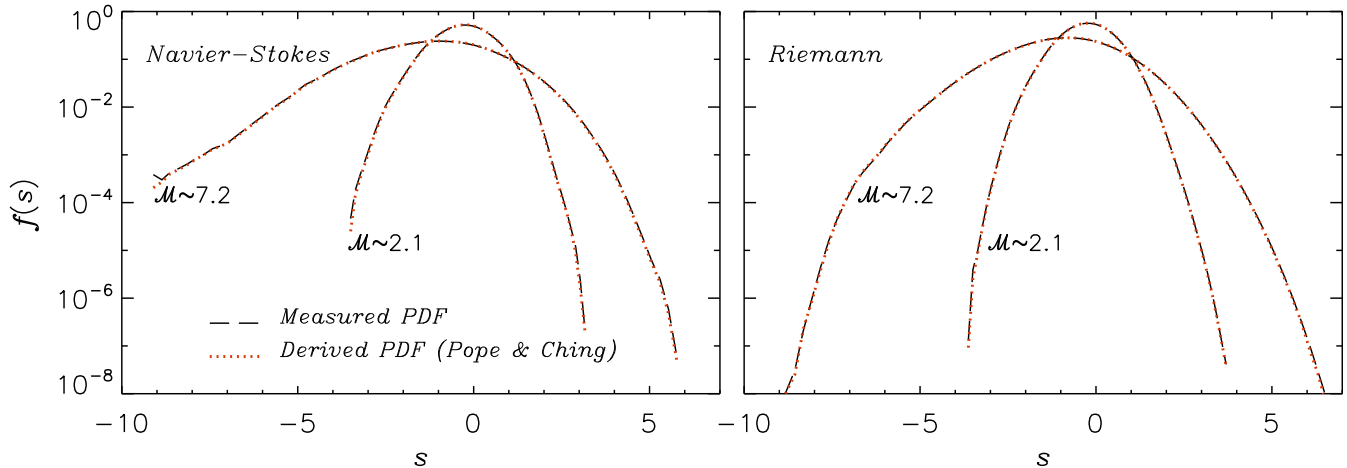}
\caption[]{Density PDFs  from 1024$^3$ N-S (left panel) and Riemann (right panel) runs at $\mathcal{M} =2.1$ and 7.2.  
The long-dashed lines show the directly measured PDFs, while the dotted lines correspond to the PDFs derived from the Pope \& Ching formula, 
Equation (\ref{stimesolution2-eq}). 
}
\label{pdf: popeching}
\end{figure}

\section{B. Deriving Equation (7) Using the Pope \& Ching Formulation} 
We give an alternative derivation for Equation (\ref{solution}) in \S 2.1.
The derivation first applies the formulation of Pope \& Ching (1993)  in Appendix A to the density PDF in the Lagrangian frame, and then obtains the PDF in the Eulerian frame 
from its relation with the Lagrangian PDF. 
We start with a number of definitions concerning the PDF in the Lagrangian frame.  We first 
consider a fluid element initially located at, ${\bs x}_0$, at time 0, and define its fine-grained PDF as 
$g_L(\zeta; {\bs x}_0, t) =  \delta(\zeta - s_L({\bs x}_0, t))$, where $s_L({\bs x}_0, t)$ is the logarithm of the density of the fluid element at time $t$.  For any flow quantity,  $\psi$, we denote it as  $\psi({\bs x}, t)$ and $\psi_L({\bs x}_0, t)$, respectively, in the Eulerian and Lagrangian frames.
We define the Lagrangian average, $\langle \psi_L \rangle_L$, 
as the density-weighted average over all fluid elements at time 0, i.e.,  $ \langle \psi_L \rangle_L= \frac{1}{M}\int_V \psi_L({\bs x}_0, t) \rho_0({\bs x}_0) d^3x_0$, with $V$ 
and $M$ are the total volume and total mass of the flow, and $\rho_0({\bs x}_0)$ the density field at time 0.  The factor, $\rho_0({\bs x}_0)$, gives more weight to the dense regions at time zero, 
where there are more fluid elements. The coarse-grained Lagrangian PDF  is then defined as  $f_L(\zeta; t) \equiv \langle g_L \rangle_L =  \langle \delta(\zeta - s_L({\bs x}_0, t)) \rangle_L$.
For any flow quantity, $\phi$, the Lagrangian conditional mean is given by ${\langle \phi_L |s_L=\zeta \rangle_L } = \langle \phi_L g_L \rangle_L/f_L$. 

With these definitions, one may obtain the Lagrangian PDF by the same derivation similar as in Appendix A,  
\begin{equation}
f_L(\zeta) = \frac{C_L}{\langle {\dot{s}_L}^2|s_L=\zeta \rangle_L } \exp\left(\int\limits_{0}^{\zeta} \frac{\langle {\ddot{s}_L} |s_L=\zeta' \rangle_L} {\langle  {\dot{s}_L}^2|s_L=\zeta' \rangle_L} d\zeta'  \right),
\label{LagrangianPDF-eq}
\end{equation}
where $\dot{s}_L = d s_L({\bs x}_0, t)/dt$ and $\ddot{s}_L = d^2s_L({\bs x}_0, t)/dt^2$ are the first and second Lagrangian time derivatives of $s$.  

We now show that the Lagrangian PDF is related to the Eulerian PDF ($f(\zeta)$)
as $f_L (\zeta)= \exp(\zeta) f(\zeta)$. The definition of the Lagrangian PDF is $f_L(\zeta; t) = \frac{1}{M}\int_V \delta(\zeta - s_L({\bs x}_0, t))  \rho_0({\bs x}_0) d^3x_0$,  and 
a variable change from the initial position, ${\bs x}_0$, of a fluid element to its position, ${\bs x}$, at $t$ gives $f_L(\zeta; t) = \frac{1}{M}\int_V \delta(\zeta - s({\bs x}, t))  \rho({\bs x}, t) d^3x$ where we have used the mass conservation $\rho({\bs x}, t) d^3x = \rho_0({\bs x}_0) d^3x_0$.  Since $s \equiv \ln(\rho/\bar {\rho}) $, it follows that 
$f_L(\zeta; t) = \frac{1}{V}\int_V \delta(\zeta - s({\bs x}, t)) \exp(s({\bs x}, t))d^3x$, which is equal to $\exp(\zeta) f(\zeta)$, as claimed.

In a similar way, one can show that the Lagrangian conditional mean of any flow quantity, $\phi$, is identical to its Eulerian conditional mean, i.e., $\langle \phi_L |s_L=\zeta \rangle_L = \langle \phi |s=\zeta \rangle$.  Finally, from the continuity equation, we have $\dot{s}_L= -\nabla \cdot {\bs u}$ and $\ddot{s}_L= -d(\nabla \cdot {\bs u})/dt$, 
and inserting them into Equation (\ref{LagrangianPDF-eq}) gives Equation (\ref{solution}) in the text.

\section{C. Verifying the Pope \& Ching Formula} 

The formula, Equation (\ref{stimesolution2-eq}), of Pope \& Ching (1993) is a general, mathematical result, and it applies to any quantity in any 
statistically stationary system.  In order to test the Pope-Ching formula for the density PDF in supersonic turbulence, we compare in Figure \ref{pdf: popeching} the PDFs 
calculated from Equation (\ref{stimesolution2-eq}) against those directly measured from simulation data. The time derivatives of $s$ needed in Equation (\ref{stimesolution2-eq}) are 
obtained by computing the changes of $s$ between neighboring time steps as the simulation was running. 
The dotted lines in Figure \ref{pdf: popeching} show the PDFs derived from the Pope \& Ching formula for simulated flows at $\mathcal{M} =2.1$ and $7.2$. 
For both the N-S and Riemann runs, the dotted lines are almost indistinguishable from the directly measured PDFs (long-dashed lines). 
The same is found for all Mach numbers, confirming the validity of the Pope \& Ching formula. The agreement between the derived PDFs and the measured ones also suggests that the 
time derivatives of $s$ used in the formula are accurate. In fact, the time derivatives were directly computed from the temporal changes of $s$ in the simulated 
flows, and thus correctly represent the evolution of the density field.  

The Pope \& Ching formula is essentially a mathematical result as it 
ignores what drives the evolution of the quantity of interest. Therefore, it does not help reveal the physics behind the PDF, 
unless the expectations of the time derivatives, $\dot{ s}$ and $\ddot{ s}$, are related to the hydrodynamic equations. 
The agreement of the PDFs derived from the Pope-Ching formula with the measured PDFs is not 
by itself a proof for the accuracy of the density PDFs measured in the Riemann simulations. 
However, the Pope-Ching formula can be used to understand the reliability of the PDF in Riemann runs (see \S 5) 
by examining whether the conditional expectations of $\dot{ s}$ and $\ddot{ s}$ are free from the numerical artifacts 
that affect the spatial derivatives (see Appendix D).

\begin{figure}
\begin{center}
\includegraphics[width=0.5\columnwidth]{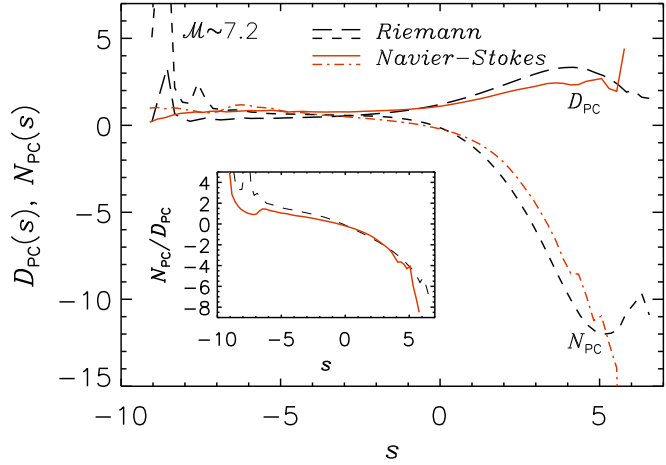}
\caption[]{Conditional averages of the first and second time derivatives of $s$ in the N-S (red) 
and Riemann (black) runs at Mach 7.2.  $D_{\rm PC}(s) $
and $N_{\rm PC}(s) $ are defined as ${\langle \dot{ s}^2 |s\rangle}/\langle \dot{s}^2 \rangle$ and 
$\langle \ddot{s}|s \rangle/\langle \dot{s}^2 \rangle$, respectively. }
\label{pdf: timederivatives}
\end{center}
\end{figure}  

\section{D. Conditional mean time derivatives in the Navier-Stokes and Riemann Runs}

The formula, Equation (\ref{stimesolution2-eq}), of Pope \& Ching, suggests that the PDF of the density may be determined
by the conditional averages of the first and second order time derivatives of $s$. If the conditional mean time derivatives 
in two simulated flows are the same, then the density PDF would also be the 
same. 
In \S 5, we have shown that, at the same Mach number, the density PDFs from the N-S run and Riemann run are 
consistent, despite substantial differences between the two runs. 
Based on the Pope-Ching formula, one way to understand the agreement of the PDFs in the two types of simulations 
is to compare the conditional averages of the time-derivatives of $s$.

Figure \ref{pdf: timederivatives} plots that $D_{\rm PC}(s)$ and $N_{\rm PC}(s)$ measured from the N-S (red) and Riemann (black) 
runs at ${\mathcal M} =7.2$.  $D_{\rm PC}$ and $N_{\rm PC}$ were defined as the conditional averages of 
$\dot{s}$ and $\ddot{s}$, normalized to $\langle \dot{s}^2 \rangle$. 
We see that $D_{\rm PC}(s)$ and $N_{\rm PC}(s)$  from the two runs show 
exhibit similar behaviors with $s$. The inset shows  the ratio of $N_{\rm pc}(s)$ to $D_{\rm pc}(s)$, 
which appears as the exponent in the exponential factor in the Pope \& Ching formula (Eq.\ \ref{DNstimesolution-eq}). 
The ratio from the N-S and Riemann runs agree with each other quite well, except at extreme densities. 
Similar behaviors are also found at ${\mathcal M} =2.1$ and $3.8$.  
To conclude, even though the N-S and Riemann runs were carried out very differently, the evolution of $s$ at a given 
density level is statistically similar, and thus the density PDFs in the two types of simulations runs are consistent with each other, 
as found in \S 5.

\end{document}